\documentclass[aps,prd,amsmath,amssymb,12pt]{revtex4}
\usepackage{graphicx}
\usepackage{bm}
\usepackage{tcolorbox}
\newcommand{\be}{\begin{equation}}
\newcommand{\ee}{\end{equation}}
\newcommand{\bea}{\begin{eqnarray}}
\newcommand{\eea}{\end{eqnarray}}
\newcommand{\hf}{\frac12} 
\newcommand{\nn}{\cr}
\def\journal#1#2#3#4{{#1} {\bf #2}, #3 (#4)}
\def\eq#1{(\ref{#1})}
\def\la{\langle}
\def\ra{\rangle}
\def\Tr{{\mathrm{Tr}}}
\def\ord#1{{\cal O}(#1)}
\def\mr#1{{\mathrm{#1}}}
\def\v#1{{\bm{#1}}}
\def\dv#1{\dot{\v{#1}}}

\def\dt{{\Delta t}}
\def\hx{\hat x}
\def\hy{\hat y}
\def\hv#1{\hat{\v{#1}}}
\def\ih{\frac{i}{\hbar}}

\begin{document}
\title{Electric current in a translation invariant environment}
\author{Janos Polonyi, In\`es Rachid}
\affiliation{Strasbourg University, CNRS-IPHC,23 rue du Loess, BP28 67037 Strasbourg Cedex 2 France}
\date{\today}

\begin{abstract}
The master equation for the reduced density matrix of a charged particle interacting with a translation invariant weakly coupled environment is considered. The electric current is renormalized by the system-environment interaction, leading to a direct signature of the environment in the bremsstrahlung. The general solution is given in the absence of the external electromagnetic field and the spread and the decoherence of a wave packet are followed. The increased complexity and importance of the boundary conditions for the density matrix are pointed out.
\end{abstract}
\date{\today}
\maketitle

\section{Introduction}
While we usually follow only a few appropriately chosen degrees of freedom in our experiments, the theoretical background is canonical quantization, developed for closed many-body dynamics with prescribed initial and final states. The effective theories are supposed to fill the gap, where the environment degrees of freedom are eliminated dynamically and their impact on the observed system is represented by the highly involved effective system dynamics. 

There are two different ways the effective dynamics can be established, by working in the operator or the path integral formalism. The projector operator method \cite{nakajima,zwanzig,argyres,grabert} leads to an effective equation of motion for the reduced system density matrix, the master equation. The master equation has already been obtained for a test particle \cite{zeh} by including decoherence and then taking the dissipation into account followed shortly \cite{gallis,diosi,adler}. The more systematic description based on kinetic theory, the Born approximation scheme \cite{vacchini,lanz,vacchinie}, and the inclusion of the higher-order perturbative contributions within the gas \cite{hornbergerk} was added later, together with the traditional many-body treatment of the environment \cite{dodd}. The master equation, derived within a simple but generic harmonic model \cite{agarwal,calderialeggett,unruh}, offers a powerful phenomenological treatment, applicable for weakly interacting environment.

The direct path integral expression for reduced system density matrix \cite{feynman} can be obtained within the closed time path (CTP) formalism, introduced in quantum field theory \cite{schw}, condensed matter physics \cite{keldysh,kamenev,rammer} and used later in an ever-widening area \cite{calzetta}. The exact master equation with memory for the reduced density matrix was derived in this formalism for harmonic effective dynamics in ref. \cite{hu9293} and the master equation employed below represents a Markovian truncation. The time evolution of one and two Gaussian wave packets has been investigated in refs. \cite{joos,paz}, respectively.

The two schemes are usually applied in a complementary manner: While the operator approach reproduces directly the density matrix, summarising our information about the system, the path integral formalism is powerful to generate the Green functions of the coordinate operator. The present work grew out from an attempt to apply a simple master equation for the introductory problems to quantum mechanics, namely for the spread of the wave packet and the propagation in the presence of a step function potential. 

The solution of the master equation provides a simple and efficient way to deal with the quantum Brownian motion, demonstrated here by the treatment of the spread of a Gaussian wave packet. A clear fingerprint of the dissipative forces is found in the electric current of the wave packet. One encounters two unexpected features of open dynamics, as well. One is related to the incompleteness of the effective dynamics by deriving the master equation, namely the missing auxiliary conditions to assure the self-adjointness of the momentum operator. Another surprise is the renormalization of the electric current by dissipative processes.

Some unusual kinematical features of the mixed state, present already for closed dynamics, are collected in Section \ref{mixeds}. The master equation is introduced in section \ref{masters}, its derivation is based on the most general, translation invariant, harmonic, local $\ord{\partial_t^2}$ interaction Lagrangian with an environment and follows the strategy of finding the Fokker-Planck equation. An important new element of the interactions the modification of the electric current is shown in section \ref{ecurrs}. The solution of the master equation is presented in section \ref{tiecs}, followed in section \ref{decs} by the demonstration of the instantaneous and the dynamical decoherence for the linear superposition of two plane waves. The spread and the decoherence of a wave packet is worked out in section \ref{wavepacketss}. We collected few remarks in section \ref{boundcs} about the role of the spatial boundary conditions for the solutions of the master equation. The main results are summarized in section \ref{concls} and an Appendix is added for the sake of completeness about the effective equation of motion in the CTP formalism.

\section{Closed dynamics of mixed states}\label{mixeds}
We start with a particle whose closed dynamics is defined by the Schr\"odinger equation, $i\hbar\partial_t\psi=H\psi$. The first kinematical step toward the handling of open interaction channels, the main topics of this work, is to extend the closed dynamics for mixed states. One introduces at this point the Neumann equation $\partial_t\rho=-i[H,\rho]/\hbar$ for the density matrix $\rho$. It is worth commenting on the apparence of a few novel points of this equation of motion compared to the Schr\"odinger equation.

\subsection{Linearity} 
One can write the Neumann equation in the form ${\cal L}_0\rho=0$  where the Liouville operator ${\cal L}_0$ acts on the linear space of operators $\cal A$ corresponding to the Hilbert space $\cal H$ of pure states, called Liouville space. Since the Neumann equation acts linearly on the Liouville space the linear superposition of solutions remains a solution as in the case of the Schr\"odinger equation. However, this similarity hides a fundamental difference between the wave function and the density matrix as far as the physical interpretation is concerned, namely, the expectation values are quadratic, $\la\psi|A|\psi\ra$, or linear, $\Tr[\rho A]$, in terms of the wave function or the density matrix, respectively. As a result, there is {\em no interference} between the terms of the linear superposition in the expectation values, $\Tr[A(\rho+\rho')]=\Tr[A\rho]+\Tr[A\rho']$. 

The linearity of the Schr\"odinger equation allows us to construct new solutions by adding up solutions by conserving the norm, $\la\psi|\psi\ra$, during the time evolution. The same holds for the Neumann equation where the total probability, $\Tr[\rho]$, is conserved. The circumstance that the conserved quantity is linear in the density matrix suggests the splitting of the solutions into two classes, (i) well-defined total probability, $0<|\Tr[\rho]|<\infty$ and (ii) ill-defined  total probability, $\Tr[\rho]=0$ or $\Tr[\rho]=\infty$. The time dependence of solutions of class (ii) is not restricted by the conservation of the total probability. There are no such unrestricted components in the solution of the Schr\"odinger equation where each component of the wave function is restricted by the unitarity of the time evolution.

\subsection{Auxiliary conditions} 
One needs auxiliary conditions to make the solution of the Neumann equation unique. Let us consider the one-dimensional case for the sake of simplicity where the solution can be fixed uniquely by providing $\psi(t,x_0)$ and $\nabla\psi(t,x_0)$ for some fixed $x_0$ and for each time $t$. The solution of the Neumann equation, a second-order hyperbolic equation, can be fixed uniquely by prescribing the density matrix $\rho(x_+,x_-)$ and its normal derivative along the boundary of some region on the plane $(x_+,x_-)$ in a time-dependent manner. There are pathological regions, bounded by characteristic lines, where the two functions of the boundary conditions can not be freely chosen. The characteristic curves of Neumann equation are the straight lines $x_+\pm x_-=$const.. Thus, the presence of the mixed state contributions increases dramatically the amount of additional information we have to provide as auxiliary conditions to identify a unique solution compared to the pure state.

\subsection{Matching conditions} 
Let us now assume that there is a potential in the Hamiltonian $U(x)=\lambda\delta(x)$ up to terms with finite discontinuity at $x=0$. The continuity of the wave function at $x=0$ is usually taken for granted in closed systems. Apart from a number of incomplete arguments, the most convincing way to derive this condition for the step function potential is to replace the Heaviside function by a linearly rising function within the interval $0\le x\le\eta$ and showing by the help of the analytic structure of Airy's function that the stationary wave function and its first derivative remain continuous in the limit $\eta\to0$  \cite{branson}. It is easy to check that this argument remains valid for the Neumann equation, as well. 

The integration of the Schr\"odinger equation across the singularity yields 
\be
\mr{Disc}_x\nabla\psi(t,0)=\frac{2mg}{\hbar^2}\psi(t,0)
\ee
where $\mr{Disc}_zf$ denotes the discontinuity of the function $f$ in the variable $z$. A similar procedure leads to 
\bea\label{matcondn}
\mr{Disc}_x\nabla_+\rho(t,0,x_-)&=&\frac{2mg}{\hbar^2}\rho(t,0,x_-),~~~x_-\ne0,\nn
\mr{Disc}_x\nabla_-\rho(t,x_+,0)&=&\frac{2mg}{\hbar^2}\rho(t,x_+,0),~~~x_+\ne0.
\eea
The matching conditions, imposed on a factorizable density matrix, $\rho(x_+,x_-)=\psi(x_+)\psi^*(x_-)$, contain the same information at any points along the lines $x_+=0$ and $x_-=0$ since the trivial multiplicative factor, depending on the other variable, $x_-$ or $x_+$, respectively, drops out. The matching conditions for the Neumann equation spreads over lines to accommodate the richness of the mixed states.

The Neumann equation is replaced by a master equation in the rest of this work. What is the importance of the above-mentioned peculiarities of the closed dynamics of mixed states in the presence of open interaction channels? Recall that the mixed states arise from two different physical considerations. (a) We have some missing classical information about the quantum state and all we know is that with  probability $p_n$ the system is in the normalized state $|\psi_n\ra$. The expectation value of observables can be obtained by the help of the density matrix $\rho=\sum_n|\psi_n\ra p_n\la\psi_n|$. (b) One splits a full closed system into the observed subsystem and its environment. The Hilbert space of pure states form the direct product, ${\cal H}_{tot}={\cal H}_{obs}\otimes{\cal H}_{env}$ and the expectation value of observables of the observed subsystem is reproduced by the reduced density matrix, $\rho=\Tr_{env}[\rho_{tot}]$, by taking the trace of the full density matrix over the environment Hilbert space. The nonfactorizability of the density matrix arises from the the nontrivial probability assignement of the pure state and from the observed subsystem-environment entanglement in case (a) and (b), respectively. While case (a) appears already for closed dynamics, the case (b) belongs to open dynamics.

The Neumann equation is linear in case (a), and each component of the density matrix is of class (i). The linearity of the master equation in case (b) requires the factorization of the initial density matrix into the product where the factors correspond to the observed subsystem and to its environment. When this is granted, then solutions of class (ii) appear and represent decoherence,  cf. Sections \ref{decs} and  \ref{wavepacketss}. The decoherence is a dynamical process and naturally occurs even if the initial density matrix is nonfactorizable only its manifestation is more involved. The increased richness of the mixed states represents the classical probability distribution of the pure states in closed dynamics and encodes the system-environment entanglement in open dynamics. The proper handling of this richness requires more care in defining the Liouville space, to be addressed in Section \ref{boundcs}.

\section{Effective equation of motion}\label{masters}
We consider here a particle propagating on a fixed electromagnetic field and coupled weakly to its macroscopic environment. The macroscopic environment is supposed to be homogeneous, i.e. translation and rotation invariant, and is assumed to possess low-lying excitations with sufficiently large spectral weight to generate dissipation. The closed dynamics of the particle without taking into account the interaction with the environment is defined by the Lagrangian 
\be\label{syslagr}
L_0=\frac{m_0}2\dv{x}^2(t)-e\phi(t,\v{x})+\frac{e}c\dv{x}(t)\v{A}(t,\v{x}).
\ee
where $A_\mu=(c\phi,-\v{A})$ denotes the vector potential. 

The interaction with the environment generates an open effective dynamics for the particle to be taken into account within the CTP formalism, outlined briefly in Appendix \ref{ctps}. A distinguished feature of the CTP formalism is a formal reduplication of the degrees of freedom, the use of a pair of coordinates, $\v{x}\to\hv{x}=(\v{x}_+,\v{x}_-)$ for a single particle. The need of the redoubling in the treatment of open quantum systems can be understood by noting that a mixed state is represented by the reduced density matrix $\la\v{x}_+|\rho|\v{x}_-\ra$, having twice as many variables as the wave function $\la\v{x}|\psi\ra$ of pure states. While the variables $\v{x}_\pm$ are motivated by the physical origin of the density matrix, the combinations $\v{x}=(\v{x}_++\v{x}_-)/2$, and $\v{x}_d=\v{x}_+-\v{x}_-$ are more advantageous to separate the physical coordinate ($\v{x}$), from its quantum fluctuations ($\v{x}_d$) and the density matrix is assumed to be parametrized as $\rho(\v{x},\v{x}_d)=\la\v{x}+\v{x}_d/2|\rho|\v{x}-\v{x}_d/2\ra$ below unless it is stated explicitly otherwise. Such an enrichment of the mathematical formalism allows us to cover open interactions by the variational principle and the reduced density matrix in the classical and quantum case, respectively \cite{macr}.

The reduplication of the coordinates might formally be viewed as a reduplication of the degrees of freedom. However, this is not the case, the true kinematical origin is the correlation of the quantum fluctuations in the bra and the ket sector of a mixed state of a single degree of freedom cf. Appendix \ref{ctps} and makes no harm to quantities reflecting the number of degrees of freedom such as thermodynamical potentials. In a similar manner the symmetries are not to be redoubled in open dynamics \cite{elementary}, a principle which becomes important in dealing with gauge theories. In the present context, we treat the electromagnetic field as an external, nondynamical environment of a charged particle and do not redouble it. 

The equation of motion for the reduced density matrix, called the master equation, will be derived in two steps within the Markovian approximation. First, the most general effective Lagrange function is constructed, where the homogeneous system-environment interactions are represented by quadratic terms in the coordinates followed by the derivation of the master equation.

\subsection{Effective Lagrangian}
The Lagrangian \eq{syslagr} generalizes to the CTP formalism 
\be\label{syslagrctp}
L_0=\frac{m_0}2[\dv{x}^2_+(t)-\dv{x}^2_-(t)]-e\phi(t,\v{x}_+)+\frac{e}c\dv{x}_+(t)\v{A}(t,\v{x}_+)+e\phi(t,\v{x}_-)-\frac{e}c\dv{x}_-(t)\v{A}(t,\v{x}_-)
\ee
and is to be appended by additional terms representing the interactions with the environment $L=L_0+L_{infl}$, whose most general translation and rotational invariant $\ord{x^2}$ form which is local in time is 
\bea\label{inflagr}
L_{infl}&=&c_1\v{x}^2-m\omega^2\v{x}_d\v{x}+\frac{i}2d_0\v{x}^2_d\nn
&&+c_2\v{x}\dv{x}-m\nu\v{x}_d\dv{x}-m\nu_d\v{x}\dv{x}_d-im\xi\v{x}_d\dv{x}_d\nn
&&+c_3\dv{x}^2+\delta m\dv{x}\dv{x}^d+\frac{i}2d_2\dv{x}_d^2.
\eea
The terms multiplied by the coefficients $c_j$, $j=1,2,3$, can be shown to violate the Lindblad condition in the master equation and will be ignored. Furthermore, the terms with the coefficient $\nu_d$ and $m\omega^2$ generate $\v{x}$-dependent terms in the master equation and will be suppressed owing to the homogeneity assumption. The Lagrangian is, therefore, reduced to
\bea\label{efflagr}
L&=&m\dv{x}\dv{x}_d+\frac{i}2d_2\dv{x}_d^2-m\nu\v{x}_d\dv{x}-im\xi\v{x}_d\dv{x}_d+\frac{i}2d_0\v{x}^2_d-U_d+\dv{x}\v{a}_d+\dv{x}_d\v{a}
\eea
with 
\bea
\v{a}(\hv{x})&=&\frac{e}c\frac{\v{A}(\v{x}_+)+\v{A}(\v{x}_-)}2,\nn
\v{a}_d(\hv{x})&=&\frac{e}c[\v{A}(\v{x}_+)-\v{A}(\v{x}_-)],\nn
U_d(\hv{x})&=&\frac{e}c[\phi(\v{x}_+)-\phi(\v{x}_-)],
\eea
and $m=m_0+\delta m$. Note that the term proportional to $\xi$ is a total time derivative and its role is to generate a time-independent, nondynamical multiplicative factor $\exp\xi m\v{x}_d^2/2\hbar$ to the density matrix, a static decoherence ($\xi<0$) or recoherence ($\xi>0$) in the coordinate representation. The classical equation of motion for $\v{x}$ sets $\v{x}_d=0$, and the equation of motion for $\v{x}_d$ yields the Newton equation with a friction force $-m\nu\dot{\v{x}}$ and the dissipation timescale $\tau_{diss}=1/\nu$. The imaginary $d_0$- and $d_2$-dependent part of the Lagrangian controls the decoherence in the coordinate basis. The effective Lagrangian \eq{efflagr} can be derived in the case of a test particle interacting with a homogeneous gas in the leading order of the test particle and gas interaction. The results, summarized in Appendix \ref{iges}, show that the parameters $\delta m$, $\nu$, $d_0$, $d_2$ and $\xi$ are $\ord{g^2}$ expressions of the coupling constant $g$ of the test particle-gas interaction.

\subsection{Master equation}\label{maeqs}
The effective Lagrangian \eq{efflagr} yields the master equation,
\be\label{master}
0={\cal L}\rho
\ee
containing the Liouville operator
\bea\label{masterl}
{\cal L}&=&-\nabla_{Ut}+i\frac\hbar{m}\v{\nabla}_{Ax}\v{\nabla}_{Ad}+\frac{\hbar d_2}{2m^2}\Delta_{Ax}\nn
&&-\frac{d_0+d_2\nu^2-2m\nu\xi}{2\hbar}\v{x}^{d2}+\left(\frac{d_2}m\nu-\xi\right)i\v{x}_d\v{\nabla}_{Ax}-\nu\v{x}_d\v{\nabla}_{Ad}.
\eea
The partial derivatives 
\bea
\v{\nabla}_x&=&\frac\partial{\partial\v{x}}=\frac\partial{\partial\v{x}_+}+\frac\partial{\partial\v{x}_-},\nn
\v{\nabla}_d&=&\frac\partial{\partial\v{x}_d}=\hf\left(\frac\partial{\partial\v{x}_+}-\frac\partial{\partial\v{x}_-}\right),
\eea
are extended to the covariant derivatives 
\bea\label{covdd}
\nabla_{Ut}&=&\partial_t+\ih U_d,\nn
\v{\nabla}_{Ax}&=&\v{\nabla}_x-\ih\v{a}_d,\nn
\v{\nabla}_{Ad}&=&\v{\nabla}_d-\ih\v{a}_a,
\eea
by the help of the minimal coupling. The structure of the minimal coupling is better seen when the original $\v{x}_\pm$ coordinates are used:
\bea\label{masterlpm}
{\cal L}&=&-\nabla_{Ut}+i\frac\hbar{2m}(\v{\nabla}^2_{A+}-\v{\nabla}^{*2}_{A-})+\frac{\hbar d_2}{2m^2}(\v{\nabla}^2_{A+}+\v{\nabla}^{*2}_{A-}+2\v{\nabla}_{A+}\v{\nabla}^*_{A-})\nn
&&-\frac{d_0+d_2\nu^2-2m\nu\xi}{2\hbar}\v{x}^{d2}+\v{x}^d\left[\left(\frac{d_2}m\nu-\xi\right)i(\v{\nabla}_{A+}+\v{\nabla}^*_{A-})-\hf\nu(\v{\nabla}_{A+}-\v{\nabla}^*_{A-})\right]
\eea
with the covariant derivatives
\be
\v{\nabla}_{A\pm}=\frac\partial{\partial\v{x}_\pm}-\ih\v{a}(\v{x}_\pm).
\ee
The derivation of this result is presented briefly in Appendix \ref{effeoms}. The first two terms in Eq. \eq{masterl} incorporate the Neumann equation for a closed system, and the next term with $\Delta_{Ax}$ generates a decoherence-induced diffusion in the coordinate space. The $\ord{x_d^2}$ term induces Gaussian decoherence by suppressing the density matrix for large $|x_d|$. The parameter $\xi$ multiplies a total time derivative term in the effective Lagrangian \eq{efflagr} hence, it represents no dynamical process. The piece containing $\v{x}_d\v{\nabla}_{Ax}$ couples the $\v{x}$ and the $\v{x}_d$ dependence, and $-\nu\v{x}_d\v{\nabla}_{Ad}$ works against the decrease of $\rho$ in $|x_d|$.

The Liouville operator ${\cal L}$ of the master equation is linear, and the remarks in Section \ref{mixeds} apply. Note that the open dynamics increases further the set of necessary auxiliary conditions. In fact, the second derivative $\nabla^2_x$ needs the knowledge of a further function, say, $\nabla_x\rho(x_0,x_d)$, to find a unique solution.

\subsection{Generic quadratic master equation}
It is instructive to compare our master equation with the most general one-dimensional quadratic master equation of the Lindblad form  \cite{lindblad} which is
\bea\label{genmasteq}
\partial_t\rho&=&-\frac{i}\hbar\left[\frac{p^2}{2m}+\frac{m\omega^2}2x^2,\rho\right]-\frac{i}{2\hbar}(\lambda+\mu)[x,\{p,\rho\}]+\frac{i}{2\hbar}(\lambda-\mu)[p,\{x,\rho\}]\nn
&&-\frac{D_{pp}}{\hbar^2}[x,[x,\rho]]-\frac{D_{xx}}{\hbar^2}[p,[p,\rho]]-2\frac{D_{px}}{\hbar^2}[x,[p,\rho]],
\eea
\cite{sandulescu}. Our master equation \eq{master} without external field corresponds to the choice $\omega=0$:
\bea
D_{pp}&=&\frac{\hbar(d_0+d_2\nu^2-2m\nu\xi)}2,\nn
D_{xx}&=&\frac{\hbar d_2}{2m^2},\nn
D_{px}&=&\frac\hbar2\left(\frac{d_2\nu}m-\xi\right),\nn
\lambda&=&\mu=\frac\nu2,
\eea
and preserves the positivity of the density matrix for weak friction:
\be\label{lindble}
\nu^2+4\xi^2\le2\frac{d_0d_2}{m^2}.
\ee
The parametrization based on the effective Lagrangian is more speaking about their physical origin or  impact, being closer to the original full dynamics than the parameters of the master equation \eq{genmasteq}. The classical equation of motion can formally be found by canceling the imaginary part of the Lagrangian \cite{effth}, i.e., by performing the limit $d_0,d_2\to0$. While this limit is excluded by the inequality \eq{lindble}, the true classical limit induced by strong decoherence remains within reach.

\section{Electric current}\label{ecurrs}
It is easy to see that the electric current of the Schr\"odinger equation
\be\label{curr}
(n,\v{j})=\left(\psi^*\psi,\frac\hbar{2im}[\psi^*\v{\nabla}_A\psi-(\v{\nabla}_A\psi)^*\psi]\right)
\ee
is not conserved by the master equation. It is easy to check by taking the master equation at $\v{x}_d=0$ that the current
\be\label{ccurr}
(n,\v{J})=\left(\rho(\v{x},0),\frac\hbar{im}\v{\nabla}_{Ad}\rho(\v{x},0)-\frac{\hbar d_2}{2m^2}\v{\nabla}_x\rho(\v{x},0)\right).
\ee
is conserved by the open dynamics. Note that, while the gauge field $\v{A}$ appears only in the last term in the Lagrangian \eq{syslagr}, the integration \eq{expmaeq} introduces it into each space derivative of the master equation and makes the derivatives in Eq. \eq{ccurr} of electromagnetic origin. The first term of the space component represents the electric current of the Schr\"odinger equation, and the additive renormalization $\v{j}_{env}=\v{J}-\v{j}$, a reminiscent of Fick's law, is generated by the decoherence-driven diffusion and represents the polarization cloud, a background drag, induced by the test particle in its environment and makes the total probability conserved by a nonunitary time evolution. While the test particle is obviously stable, the norm of its state can be exchanged with the environment. 

The electromagnetic origin of the conserved current suggests that Eq. \eq{curr} is actually the electric current. But the renormalization of the electric current by the environment is a puzzling  point, since the conservation of the electric current is derived by the help of gauge invariance. How can a manifestly gauge invariant equation of motion violate this conservation law? The dressing of the electric current by the UV modes can be followed in QED by the method of the renormalization of composite operators, and it was believed that the electric current is protected against the cutoff scale UV renormalization by gauge invariance \cite{peskin}. However, renormalization at a finite scale can not be ruled out in such a manner. It has been realized only recently that massless photons give rise to a further tadpole counterterm at vanishing energy-momentum transfer and generate a surface term in the integral version of the continuity equation for the current \cite{collins}. Another known modification of the electric current, closer to our subject, is due to the periodic boundary conditions imposed in a finite quantization box \cite{hu}. Here, the periodic boundary conditions generate new gauge invariant degrees of freedom, the Wilson lines, which appear in the effective theory and introduce nonminimal coupling.

Let us now return to our problem by noting that the space-time derivatives, generated by the expansion \eq{purej}, appear as covariant derivatives in Eq. \eq{masterl}, allowing one to locate the conserved electric current by the help of gauge transformations
\bea
\rho(t,\hv{x})&\to&e^{i\alpha_d(t,\hv{x})}\rho(t,\hv{x}),\nn
U(t,\hv{x})&\to&U(t,\hv{x})-\hbar\partial_t\alpha(t,\hv{x}),\nn
\v{a}(t,\hv{x})&\to&\v{a}(t,\hv{x})+\hbar\v{\nabla}_d\alpha_d(t,\hv{x}),\nn
\v{a}_d(t,\hv{x})&\to&\v{a}_d(t,\hv{x})+\hbar\v{\nabla}_x\alpha_d(t,\hv{x}),
\eea
with $\alpha_d(t,\hv{x})=\alpha(t,\v{x}_+)-\alpha(t,\v{x}_-)$, $\alpha(t,\v{x})$ being an arbitrary phase transformation. The Liouville operator of the master equation transforms under gauge transformations as 
\be
{\cal L}[\v{\nabla}_{Ax},\v{\nabla}_{Ad}]\to e^{i\alpha_d}{\cal L}e^{-i\alpha_d}
={\cal L}[\v{\nabla}_{Ax}-i\v{\nabla}_s\alpha_d,\v{\nabla}_{Ad}-i\v{\nabla}_d\alpha_d].
\ee
Let us now take a density matrix $\rho$ which solves the master equation and consider the equation
\be
0=\Tr[{\cal L}[\v{\nabla}_{Ax},\v{\nabla}_{Ad}]\rho]
\ee
which can, therefore, be written as
\be
0=\Tr[e^{-i\alpha_d}{\cal L}[\v{\nabla}_{Ax}-i\v{\nabla}_s\alpha_d,\v{\nabla}_{Ad}-i\v{\nabla}_d\alpha_d]e^{i\alpha_d}\rho]
\ee
for arbitrary $\alpha$. By considering the right-hand side as a functional of the classical field $\alpha(t,\v{x})$, the Euler-Lagrange variational equation yields the continuity equation
\be
0=\left[\v{\nabla}_x\left(\frac\hbar{im}\v{\nabla}_{Ad}-\frac{\hbar d_2}{2m^2}\v{\nabla}_x\right)+\partial_t\right]\rho(\v{x},\v{0})
\ee
for infinitesimal $\alpha$ and establishes that the conserved electric current is indeed \eq{ccurr}.

The argument is reminiscent of the derivation of the Noether theorem or the Ward identities. The minimal coupling generated by the expansion \eq{expmaeq} replaces the partial space-time derivatives with covariant derivatives in the equation of motion which, in turn, contribute to the electromagnetic form factors. Among these contributions, the $\ord{x_d^0}$ terms appear in the electric current. In our case, the second term of the Lagrange function \eq{efflagr}, responsible for the velocity-dependent decoherence-generated diffusion, renormalizes the electric current. It is worthwhile noting that such a renormalization of the electric current takes place even if the environment is electrically neutral.

\section{Translation invariant elementary components}\label{tiecs}
The impact of the open interaction channels on the dynamics is sought in this section by constructing the solutions of the translation invariant master equation.

\subsection{Translations}
The translations $\v{x}_\pm\to\v{x}_\pm+\v{a}$ induce the transformation $\rho(\v{x},\v{x}_d)\to\rho(\v{x}-\v{a},\v{x}_d)$ of the density matrix and the linear subspaces ${\cal A}_\v{q}$ consisting of density matrices of the form $\rho(\v{x},\v{x}_d)=e^{i\v{q}\v{x}}\chi(\v{x}_d)$ with arbitrary $\chi(\v{x}_d)$ remain closed under translation invariant dynamics. This space is irreducible in the sense that it contains no nontrivial subspace which is left invariant by a generic translation invariant dynamics. 

This is a radical departure from the way translations act on pure states $\psi(\v{x})\to\psi(\v{x}-\v{a})$. In fact, the translations form an Abelian group whose irreducible representations within the space of pure states are one dimensional. The translation invariance of $\v{x}_d$ extends these subspaces to  ${\cal A}_\v{q}$ \cite{elementary}. The physical origin of this richness of elementary constituents is the background drag, a nontrivial polarization cloud induced by the test-particle in the environment, which follows the motion of the particle. The relevance of the subspaces ${\cal A}_\v{q}$ from the point of view of the present work is that it is sufficient to solve the master equation within them.

The solution of the master equation is sought below with the potential $e\phi(\v{x})=U_0-\v{f}\v{x}$, $\v{A}=0$ within the space ${\cal A}_q$ with an arbitrary initial condition. The translation induces a phase rotation on pure state vectors which is canceled in the density matrix. Such a cancellation leads to another interesting cancellation within the Neumann equation: The potential $U_0$ drops out, and the homogeneous force is represented by the term $i\v{f}\v{x}_d/\hbar$, leaving the Neumann equation translation invariant. This potential, being a pure gauge potential, generates an $x$-dependent phase for translations. The dropping of the constant component $U_0$ makes the same Neumann equation to cover both the propagating ($U_0<0$) and the tunneling ($U_0>0$) states. Our master equation preserves this property.

\subsection{Solution within an ${\cal A}_\v{q}$ subspace}
The solution of the master equation for the coefficient function $\chi$,
\bea\label{chieq}
\partial_t\chi(t,\v{x}_d)&=&-\left(\frac{\hbar\v{q}}m+\v{x}_d\nu\right)\v{\nabla}_d\chi(t,\v{x}_d)\nn
&&+\left[-\frac{\hbar\v{q}^2d_2}{2m^2}+\left(i\frac{\v{f}}\hbar-\frac{\v{q}d_2}m\nu+\v{q}\xi\right)\v{x}_d-\frac{d_0+d_2\nu^2-2m\nu\xi}{2\hbar}\v{x}_d^2\right]\chi(t,\v{x}_d),
\eea
is the easiest to find by first constructing the characteristic curve, a trajectory $\bar x_d(t)$, satisfying the equation of motion
\be
\partial_t\bar{\v{x}}_d(t)=\nu\bar{\v{x}}_d(t)+\frac{\hbar\v{q}}m,
\ee
together with the initial condition $\bar{\v{x}}_d(0)=\v{x}_{d0}$. The solution
\be\label{charc}
\bar{\v{x}}_d(t)=\v{x}_{d0}e^{t\nu}+\frac{\hbar\v{q}}{m\nu}(e^{\nu t}-1)
\ee
is next used to rewrite the master equation as an ordinary differential equation along the characteristic curve:
\be
\frac{d}{dt}\chi(t,\bar{\v{x}}_d(t))=\left[\left(i\frac{\v{f}}\hbar-\frac{\v{q}d_2}m\nu+\v{q}\xi\right)\bar{\v{x}}_d(t)-\frac{d_0+d_2\nu^2-2m\nu\xi}{2\hbar}\bar{\v{x}}_d^2(t)-\frac{\hbar\v{q}^2d_2}{2m^2}\right]\chi(t,\bar{\v{x}}_d(t)).
\ee
This is easy to solve with the initial condition $\chi(0,\v{x}_d)=\chi_i(\v{x}_d)$:
\be\label{gsol}
\rho(t,\v{x},\v{x}_d)=\chi_i\left(\v{x}_de^{-\nu t}-\frac{\hbar\v{q}}{m\nu}(1-e^{-\nu t})\right)e^{a_\v{q}(t)+\Omega_\v{q}t+i\v{q}\v{x}+i\v{k}_\v{q}(t)\v{x}_d-\frac{d(t)}2\v{x}_d^2},
\ee
where
\bea\label{gsolpar}
a_\v{q}(t)&=&i\frac{\v{q}\v{k}_f\hbar}{m\nu}(1-e^{-\nu t})+\frac{\hbar\v{q}^2}{4m^2\nu^3}[d_0(e^{-2\nu t}-4e^{-\nu t}+3)-d_2\nu^2(1-e^{-2\nu t})]\nn
&&-\frac{\hbar\v{q}^2\xi}{2m\nu^2}(1-2e^{-\nu t}+e^{-2\nu t}),\nn
\Omega_\v{q}&=&-\frac{\v{q}\hbar}m\left(i\v{k}_f+\frac{d_0\v{q}}{2m\nu^2}\right),\nn
\v{k}_\v{q}(t)&=&(1-e^{-\nu t})\left(\v{k}_f-i\v{q}\frac\xi\nu e^{-\nu t}\right)-i\frac{\v{q}}{2m\nu^2}[d_0(1-e^{-\nu t})^2-d_2\nu^2(1-e^{-2\nu t})],\nn
d(t)&=&\frac{1-e^{-2\nu t}}{\ell^2_{dec}}.
\eea
The wave number $\v{k}_f=\v{f}/\hbar\nu$ corresponds to the classical drift velocity $\v{v}_f=\hbar\v{k}_f/m=\v{f}/m\nu$, and
\be\label{decl}
\frac1{\ell^2_{dec}}=\frac{d_0+d_2\nu^2}{2\hbar\nu}-\frac{m\xi}{\hbar}
\ee
denotes the asymptotic decoherence length scale. The dissipation timescale $\tau_{diss}=1/\nu$ gives the onset of the decoherence, too. The linear superposition of the solutions with an initial condition of the form $\chi_i(\v{x}_d)=e^{i\v{x}_d\v{k}}$ allows us to construct the time dependence for any initial density matrix obtained from the Hilbert space of square integrable wave functions.

The solutions \eq{gsol} can easily be extended. It is easy to see that the transformation 
\be\label{oshift}
\rho\to\left(1+\frac{m\nu}\hbar\frac{\v{q}\v{x}_d}{\v{q}^2}\right)^{-\frac{\Delta\Omega}\nu}\rho
\ee 
leads to the shift, $\Omega_\v{q}\to\Omega_\v{q}+\Delta\Omega$, without modifying the other parameters of the solution. Thus the asymptotic frequency spectrum of the master equation is infinitely degenerate. Such a high degree of degeneracy is a hallmark of dissipation, an invisible gapless environment. Another indication of dissipative forces, memory loss, is clearly visible in the prefactor of the exponential function in Eq. \eq{gsol}, which determines the long time asymptotic solution only by the behavior of $\chi_i(\v{x}_d)$ in a small environment of $\v{x}_d=\hbar\v{q}/m\nu$. Lowering of the frequency of the solution \eq{gsol} by $\Delta\Omega=-n\nu$ with $n$ being a positive integer the density matrix receives an $n$th-order polynomial of $\v{x}_d$ as a multiplicative factor. 

Another infinite set of solutions can be obtained by noting that the wave vector $\v{q}$ is arbitrary. Thus the transformation  $\rho\to i\v{\nabla}_\v{q}\rho$ yields another solution of the linear master equation. This amounts to the family of nonhomogeneous solutions
\be
\rho_P=P(i\v{\nabla}_\v{q})\rho
\ee
where $P(\v{x})$ stands for a polynomial of finite order.

\subsection{Infinitesimal environment interaction}
It is instructive to inspect the infinitesimal system-environment coupling limit, $g\to0$ for $\v{f}=0$. The dependence of the effective dynamics on the coupling constant $g$ can be made explicit by performing the change $V\to g^2V$ in Eq. \eq{effpar}, leading to $d_0=g^2\tilde d_0+\ord{g^4}$, $d_2=g^2\tilde d_2+\ord{g^4}$, $\nu=g^2\tilde\nu+\ord{g^4}$, and, $\xi=g^2\tilde\xi+\ord{g^4}$ where the ignored $\ord{g^4}$ terms come from higher loop contributions in eliminating the environment. 

The lack of a gap in the environment spectrum can be established only after an infinitely long time observation \cite{irr}, leading to the noncommutativity of the limits $t\to\infty$ and $g\to0$: On the one hand, the solution of the time-dependent master equation is a continuous function of $g$, and the time evolution of the closed dynamics is recovered as $g\to0$ at any finite time. On the other hand, the relaxed stationary solution is influenced by an infinitesimal interaction and develops a noncontinuous limit at $g=0$. In other words, the dynamics converges in a nonuniform manner in the limit of vanishing system-environment coupling. The structure recovered in this limit is universal, as are the laws of statistical mechanics, i.e., remains qualitatively the same for weakly open systems, independently of the details of the infinitesimal system-environment interactions.

Let us consider the unique relaxed state which is homogeneous, $\v{q}=0$. One would expect that the environment decouples and the closed dynamics is recovered as $g\to0$: however, one finds terms containing the ratio of the type $\ord{g^2}/\ord{g^2}=\ord{g^0}$ for $t\gg1/g^2\tilde\nu$, meaning that an arbitrarely weak environment interaction may have a finite impact on the relaxed state. In particular, the decoherence length approaches the nontrivial value $\ell_{dec}^*=\sqrt{2\hbar\tilde\nu/\tilde d_0}$. The discontinuity of the relaxed state when $g\to0$ indicates that no isolation can prevent a particle from developing a universal $\ord{g^0}$ Gaussian decoherence:
\be
\rho=e^{-\frac{\v{x}^2_d}{2\ell^{*2}_{dec}}}[1+\ord{g^2}],
\ee
rendering the application of closed dynamics an unrealistic approximation for the long time dynamics of real, propagating particles. It is worthwhile noting that the relaxed translation invariant density matrix is the Gibbs operator with a quantum-fluctuations-induced temperature, $T_q=\hbar^2/m\ell^{*2}_{dec}k_B$ \cite{gaussian}. The  discontinuity as $g\to0$ does not imply singular dynamics: the time needed to establish the universal decohered state  diverges as $g\to0$. The solutions with $\v{q}\ne0$ display singular $\v{x}_d$ dependence in the limit $g\to0$, but they are suppressed for $t\gg m^2\tilde\nu^2g^2/\hbar\tilde d_0\v{q}^2$ due to decoherence.

\section{Instantaneous and dynamical decoherence}\label{decs}
The decoherence is a dynamical process, it builds up during the time evolution, and one can characterize from  different points of view \cite{dyndec}. The suppression of the off-diagonal density matrix elements is basis dependent and can be considered in the coordinate or the momentum basis. The simplest way to measure the instantaneous strength of the decoherence is to take the quantum state at a fixed time, to consider the decrease of its matrix elements with increasing off-diagonality, given by $\v{x}_d$ in the coordinate basis, and to identify the characteristic length scale of the decrease. The time evolution of an initial pure plane wave state $\psi_i(\v{x})=e^{i\v{k}_i\v{x}}$,
\be
\rho(t,\v{x},\v{x}_d)=e^{i[\v{k}_ie^{-\nu t}+\frac{\v{f}}{\hbar\nu}(1-e^{-\nu t})]\v{x}_d-\frac{1-e^{-2\nu t}}{2\ell_{dec}^2}\v{x}^{d2}},
\ee
yields the instantaneous coordinate decoherence length scale
\be
\ell_{inst,dec}(t)=\frac{\ell_{dec}}{\sqrt{1-e^{-2\nu t}}}.
\ee
The stationary Wigner function at $t\to\infty$,
\bea
w(\v{x},\v{k})&=&\int d^3x_de^{-i\v{k}\v{x}_d}\rho(\v{x},\v{x}_d)\nn
&=&e^{-\frac{\ell_{dec}^2}2(\v{k}-\v{k}_f)^2}
\eea
reveals that a particle, dragged by a homogeneous force, relaxes to a completely decohered wave packet with Gaussian momentum distribution, centered at the drift momentum and spread to the inverse asymptotic decoherence length. 

The decoherence in momentum space is important to understand the suppression of the interference between macroscopically different states. In such an inquiry, one chooses a pure initial state as a linear superposition of two different system states and an observable and looks for dynamical decoherence, which manifests through the suppression of the interference contribution to the observable. Let us choose the initial state which is the linear superposition of two plane waves with wave vectors $\v{k}_1$ and $\v{k}_2$, $\psi_i(\v{x})=c_1e^{i\v{k}_1\v{x}}+c_2e^{i\v{k}_2\v{x}}$. 

The density matrix evolves in time as
\bea\label{twopw}
\rho(t,\v{x},\v{x}_d)&=&e^{-\frac{1-e^{-2\nu t}}{2\ell_{dec}^2}\v{x}_d^2}\{|c_1|^2e^{i[\v{k}_1e^{-\nu t}+\v{k}_f(1-e^{-\nu t})]\v{x}_d}+|c_2|^2e^{i(\v{k}_2e^{-\nu t}+\v{k}_f(1-e^{-\nu t}))\v{x}_d}\nn
&&+c_1c_2^*e^{i(\v{k}_1-\v{k}_2)\v{x}+\frac{i}2(\v{k}_1+\v{k}_2)[\v{x}_de^{-\nu t}-\frac{(\v{k}_1-\v{k}_2)\hbar}{2m\nu}(1-e^{-\nu t})]+a_{\v{k}_1-\v{k}_2}+\Omega_{\v{k}_1-\v{k}_2}t+i\v{k}_{\v{k}_1-\v{k}_2}x_d}\nn
&&+c^*_1c_2e^{i(\v{k}_2-\v{k}_1)x+\frac{i}2(\v{k}_1+\v{k}_2)[\v{x}_de^{-\nu t}-\frac{(\v{k}_2-\v{k}_1)\hbar}{2m\nu}(1-e^{-\nu t})]+a_{\v{k}_2-\v{k}_1}+\Omega_{\v{k}_2-\v{k}_1}t+i\v{k}_{\v{k}_2-\v{k}_1}\v{x}_d}]\}.
\eea
The first, common multiplicative factor describes instantaneous decoherence in the coordinate space and the two terms in the first line belong to the diagonal contribution in the momentum basis. The next two lines contain the off-diagonal contribution of the momentum basis whose weight decreases exponentially in time with the dynamical decoherence timescale
\be
\tau_{dyn,dec}=-\frac1{\mr{Re}\Omega_{\v{k}_2-\v{k}_1}}=\frac{2m^2\nu^2}{d_0\hbar(\v{k}_2-\v{k}_1)^2}.
\ee
One finds that the decoherence builds up faster at larger off-diagonality in momentum space. The decrease of the weight in Rq. \eq{twopw} with the increase of the off-diagonality yields the dynamical decoherence momentum scale
\be
p_{dyn,dec}(t)=\sqrt{\frac{m^2\nu^2\hbar}{d_0t}}.
\ee
The dynamical decoherence length in coordinate space \cite{dyndec}, 
\be
\ell_{dyn,dec}(t)=\sqrt{\frac{\hbar}{d_0t}},
\ee
suggests the existence of a basis-independent dynamical decoherence mechanism, $p_{dyn,dec}(t)/m=\nu\ell_{dyn,dec}(t)$.

\section{Wave packets}\label{wavepacketss}
The simplest generalization of the case of few plane waves is the propagation of a wave packet. This problem has been addressed by using the master equation $\nu=d_2=\xi=f=0$ \cite{joos} and by the help of a harmonic oscillator model of the environment \cite{paz}. We consider this problem by allowing all parameters of the master equation \eq{master} to be present.

We take a pure Gaussian wave packet as initial state at $t=0$:
\be\label{purewp}
\psi(\v{x})=(4\pi\sigma^2)^\frac34\int\frac{d^3k}{(2\pi)^3}e^{-\frac{\sigma^2}2(\v{k}-\v{k}_0)^2+i\v{k}\v{x}}.
\ee
The closed dynamics generates the density matrix
\be
\rho(t,\v{x},\v{x}_d)=(\pi\Delta x_0^2)^{-\frac32}e^{i\v{k}_0\v{x}_d-\frac{\v{x}^2_d}{2\ell_{eff,0}^2}-\frac{(\v{x}-\v{X}_0)^2}{\Delta x_0^2}+i\kappa_0^2(\v{x}-\v{X}_0)\v{x}_d}
\ee
with time-dependent parameters. The center of the wave packet follows a free motion, $\v{X}_0=(\hbar\v{k}_0/m)t$, and the second moment assumes the well-known quadratic expression in time $\Delta x^2_0=\sigma^2+\hbar^2t^2/m^2\sigma^2$ and $\ell_{eff,0}^2=2\Delta x^2_0$. The role of the $\ord{xx_d}$ term of the exponent with $\kappa_0^2=\hbar m t/(m^2\sigma^4+\hbar^2t^2)$ can be seen in the probability flux:
\be
\v{j}=\frac\hbar{im}\v{\nabla}_d\rho(t,\v{x},0)
=\frac\hbar{m}[\v{k}_0+\kappa_0^2(\v{x}-\v{X}_0)](\pi\Delta x_0^2)^{-\frac32}e^{-\frac{(\v{x}-\v{X}_0)^2}{\Delta x_0^2}}.
\ee
The first term describes the homogeneous motion of the center and the spread of the wave packet is reflected in the second term, a comoving flux pointing outward from the center of the wave packet.

The initial density matrix
\be
\rho_i(\v{x},\v{x}_d)=(4\pi\sigma^2)^\frac32\int\frac{d^3k_+d^3k_-}{(2\pi)^6}e^{-\frac{\sigma^2}2[(\v{k}_+-\v{k}_0)^2+(\v{k}_--\v{k}_0)^2]+i(\v{k}_+-\v{k}_-)\v{x}+i\frac{\v{k}_++\v{k}_-}2\v{x}_d}
\ee
evolves as
\bea\label{wpdm}
\rho(t,\v{x},\v{x}_d)&=&(4\pi\sigma^2)^\frac32\int\frac{d^3k_+d^3k_-}{(2\pi)^6}\exp\biggl[-\frac{\sigma^2}2[(\v{k}_+-\v{k}_0)^2+(\v{k}_--\v{k}_0)^2]\nn
&&+a_{\v{k}_+-\v{k}_-}(t)+\Omega_{\v{k}_+-\v{k}_-}t+i\frac{\hbar\v{q}}{2m\nu}(\v{k}_++\v{k}_-)(e^{-\nu t}-1)\nn
&&+i(\v{k}_+-\v{k}_-)\v{x}+i\left(\frac{\v{k}_++\v{k}_-}2e^{-\nu t}+\v{k}_{\v{k}_+-\v{k}_-}(t)\right)\v{x_d}-\frac{d(t)}2\v{x}_d^2\biggr]
\eea
in the open dynamics, and the uniform convergence of the integrals assures positivity. The integration is straightforward and results in
\be\label{pwsol}
\rho(t,\v{x},\v{x}_d)=(\pi\Delta x^2)^{-\frac32}e^{i\v{k}_{eff}\v{x}_d-\frac{\v{x}^2_d}{2\ell^2_{eff}}-\frac{(\v{x}-\v{X})^2}{\Delta x^2}+i\kappa^2(\v{x}-\v{X})\v{x}_d},
\ee
with the time depending on the parameters
\bea
\v{X}&=&\frac{\hbar\v{k}_0}{m\nu}(1-e^{-\nu t})+\v{v}_f\left(t-\frac{1-e^{-\nu t}}\nu\right),\nn
\Delta x^2&=&\frac{N}{m^2\nu^3\sigma^2},\nn
\v{k}_{eff}&=&\v{k}_0e^{-\nu t}+\v{k}_f(1-e^{-\nu t}),\nn
\frac1{\ell^2_{eff}}&=&\frac1{2N\hbar\nu}\Biggl\{\hbar m^2\sigma^2\nu^4+d_2(1-e^{-2\nu t})\nu^3(\hbar^2+m^2\sigma^4\nu^2)\nn
&&+d_0[(1-e^{-2\nu t})m^2\sigma^4\nu^3+\hbar^2\nu(1-4e^{-\nu t}+e^{-2\nu t}(2\nu t+3)]\nn
&&+2d_0d_2(1-e^{-2\nu t})\hbar\sigma^2t\nu^3+2d_0^2d_2\hbar\sigma^2(1-e^{-\nu t})[t\nu-2+e^{-\nu t}(t\nu+2)]\nn
&&-2m\nu\xi[\hbar^2\nu(1-e^{-\mu t})^2+m^2\sigma^4\nu^3(1-e^{-2\nu t})\nn
&&+2d_0\hbar\sigma^2(1-e^{-\mu t})[t\nu-2+e^{-\mu t}(t\nu+2)]-4\xi^2\hbar m^2\sigma^2\nu^2(1-e^{-\mu t})^2]\Biggr\},\nn
\kappa^2&=&\frac{m\nu}N(1-e^{-\nu t})[d_0\sigma^2(1-e^{-\nu t})-d_2\nu^2\sigma^2(1+e^{-\nu t})+2m\xi\nu\sigma^2e^{-\mu t}+\hbar\nu e^{-\nu t}],
\eea
where
\bea
N&=&m^2\sigma^4\nu^3+\hbar^2\nu(1-e^{-\nu t})^2+\hbar\sigma^2d_0[(2\nu t-3+4e^{-\nu t}-e^{-2\nu t})+\hbar\sigma^2d_2\nu^2(1-e^{-2\nu t})]\nn
&&+2(1-e^{-\nu t})^2\hbar\sigma^2m\xi\nu.
\eea
The corresponding conserved probability flux is
\be\label{rencurrwp}
\v{J}=\frac\hbar{m}[\v{k}_{eff}+Q^2(\v{x}-\v{X})](\pi\Delta x^2)^{-\frac32}e^{-\frac{(\v{x}-\v{X})^2}{\Delta x^2}},
\ee
with 
\be\label{qnegyzet}
Q^2=\kappa^2+\frac{d_2}{m\Delta x^2}.
\ee
The limits $g\to0$ with $\v{v}_f=\tilde{\v{v}}_fg^2$, $t\to0$ and $t\to\infty$ of the parameters are detailed in Table \ref{limpar}.

\begin{table}
\caption{The parameters are given for $g\to0$, small $t\ll1$, and $t\gg1$ up to terms $\ord{g^2}$, $\ord{t^2}$ (except $Q^2$) and $\ord{e^{-\mu t}}$, respectively.}\label{limpar}
\begin{ruledtabular}
\begin{tabular}{lcccccr}
limits&$\v{X}$&$\Delta x^2$&$\v{k}_{eff}$&$\frac1{\ell^2_{eff}}$&$\kappa$&$Q^2$\\
\hline
$g\to0$&$\frac{\hbar\v{k}_0}mt$&$\sigma^2+\frac{\hbar^2t^2}{m^2\sigma^2}$&$\v{k}_0$&$\frac1{2(\sigma^2+\frac{\hbar^2t^2}{m^2\sigma^2})}$&$\frac{mt\hbar}{m^2\sigma^4+\hbar^2t^2}$&$\frac{\hbar mt}{m^2\sigma^4+\hbar^2t^2}$\\
$\nu t\ll1$&$\frac{\hbar\v{k}_0}mt$&$\sigma^2+\frac{2\hbar d_2}{m^2}t$&$\v{k}_0(1-\nu t)+\nu\v{k}_f$&$\frac1{2\sigma^2}+\frac{2\sigma^2-\ell_{dec}^2}{\ell_{dec}^2\sigma^2}\nu t$&$\frac{\hbar\sigma^2+2(m\xi-d_2\nu)}{m\sigma^4}t$&$\frac{d_2}{m\sigma^2}+\ord{t}$\\
$\nu t\gg1$&$\v{v}_f\left(t-\frac1\nu\right)$&$\frac{2\hbar d_0}{m^2\nu^2}t$&$\v{k}_f$&$\frac1{\ell^2_{dec}}$&$\frac{m(d_0-d_2\nu^2)}{2\hbar d_0}\frac1t$&$\frac{m}{2\hbar t}$
\end{tabular}
\end{ruledtabular}
\end{table}

\subsection{Irreversibility} 
The coefficients of $\v{x}$ in the exponent, a time-independent vector for a single plane wave, now depend on time owing to the interference between the pure plane waves. The gradual suppression of the $\v{x}$ dependence, the spread of the wave packet, takes place already in closed dynamics, governed by the time reversal Schr\"odinger equation and is not a genuine irreversible process: it is due to the smooth phase relations between the plane waves in the initial state. Were we able to prepare a state with highly irregular phases between the plane waves, we might observe its shrinking during the time evolution. However, the decoherence of the open dynamics is a genuine irreversible process. In fact, the master equation \eq{master} breaks time reversal invariance, because this transformation does not act on the invisible environment \cite{irr}. Hence, one expects that the spread of the wave packet will be different in open dynamics.

\subsection{Decoherence}
The Gaussian decoherence strength $1/\ell^2_{eff}$ starts with the initial value $1/2\sigma^2$, which is a relic of the spread of the pure state and becomes saturated after the full establishment of the decoherence in agreement with kinetic theory \cite{zeh}. 

\subsection{Center} 
The probability distribution $\rho(t,\v{x},\v{0})$ displays a Gaussian wave packet following the trajectory $\v{X}$. Though the relaxed motion is a trivial result of Newton's equation, its microscopical, short time details offer a more detailed picture: The motion starts with the initial wave packet velocity $\hbar\v{k}_0/m$ and a constant acceleration $\v{a}_f=\v{f}/m$, since it takes  approximately $\tau_{diss}$ delay to establish the stationary energy loss to the environment. By the time the dissipation is stabilized, the motion is delayed by $\tau_{diss}$, in a manner reminiscent of the phase shift in the scattering processes, that is to say, the rearrangement of the particle state by the external forces.

\subsection{Width} 
The spread speeds up before and slows after the dissipation and decoherence are established compared with the closed dynamics. The speedup comes from the $\ord{d_2}$ velocity-dependent decoherence term of the influence Lagrangian which directly suppresses the interference between the pure components of the wave packet with different momentum. This effect is present from the very beginning. The slowing down can be understood by recalling that the spread of the wave packet is due to the irregular phases arising from the interference between different plane wave components in the probability distribution. Both the decoherence and the friction suppress these contributions: thereby, they delay the spread. 

\subsection{Discontinuity of $g\to0$} 
On the one hand, the spread and $\ell_{eff}$ of the closed dynamics are recovered at any finite time when $g\to0$.  On the other hand, the spread is slower and the time-independent $\ell_{dec}$ is reached for infinitesimal $g$, by letting $g\to0$ after the limit $t\to\infty$. In particular, the relaxed density matrix belonging to a canonical ensemble is reached in the latter case only, by letting the weak but finite system-environment interactions act for an arbitrary long time.

\subsection{Probability flux} 
The first term in the square brackets on the right-hand side in Eq. \eq{rencurrwp} contains the wave number $\v{k}_{eff}$ reflecting the loss of the initial velocity and the classical buildup of the drift velocity. The second term of the particle flux displays flux of the amplitude $Q^2$ away from the center of the wave packet. The short time effect of the current renormalization is the emergence of a nonvanishing value of the amplitude. The long time impact of the renormalization is the reduction of the amplitude by a factor half compared to the closed dynamics. 

\subsection{Phase transformation} 
The term $i\kappa^2\v{x}\v{x}_d$ in the exponent of the density matrix performs a gauge transformation on the state, $\psi(\v{x})\to e^{i\kappa^2\v{x}^2}\psi(\v{x})$. This is not a symmetry, because the transformation acts on the state only, leaving the observables unchanged, and generates the drift velocity. 

\subsection{Charged particle} 
Let us now suppose that the test particle is charged and look into its bremsstrahlung. The induced electromagnetic field is proportional to $Q^2$, a rather involved function of the parameters of the master equation and the time. However the asymptotic values of $Q^2$ for short and long time are remarkably simple and show a clear difference compared to the closed dynamics. Rather than increasing from zero in time, $Q^2$ starts with a finite value generated by the decoherence-induced diffusion already before the dissipation builds up. After the full buildup of the dissipation, the current becomes $\nu$ independent again, and the induced electromagnetic field is reduced by a factor $1/2$ compared to the closed dynamics.

\section{Boundary conditions}\label{boundcs}
It has been mention in Section \ref{mixeds} that the Neumann equation and the master equation \eq{master} and \eq{masterl} need significantly more auxiliary conditions than the Schr\"odinger equation to encode our uncertainty about the actual pure state of the system. To underline the importance of this question we note that the issue of auxiliary conditions is problematic already in the general strategy of using the master equation to describe open interactions. In fact, there are two ways of deriving effective, open interactions for a subsystem of a full, closed system. One is the CTP path integral representation of the reduced density matrix, and it provides unique results. Another procedure is to derive the equation of motion for the reduced density matrix by making an infinitesimal time step in the effective dynamics. However, the equation of motion, a differential equation in the Markovian approximation, produces a unique solution only if some auxiliary conditions are specified. How can we recover these auxiliary conditions in the master equation approach?

There is yet another problem in using the master equation \eq{master} and \eq{masterl} which turns out to be related to the auxiliary conditions. The master equation possesses a Lindblad structure, and its solution is supposed to preserve the total probability. But the solution \eq{gsol} and \eq{gsolpar} violates this  conservation law. In fact, let consider a one-dimensional particle in the interval $0<x<L$ with the density matrix $\rho=\rho_q+\rho_{-q}$, where $\rho_q$ is a solution with the wave number $q$.  It is easy to see that the total probabilty
\be
P=2\chi_i\left(-\frac{\hbar q}{m\nu}(1-e^{\nu t})\right)e^{a_q(t)+\Omega_qt}\frac{\sin qL}q   
\ee
is time dependent. Actually, this is the interference term in Sections \ref{decs} and \ref{wavepacketss}, and decoherence consists of its decrease in time. Furthermore, the solutions with imaginary wave number $\rho_{iq}$ give an exponentially decreasing probability density in the coordinate, a throughflux of particles moving to the right with exponentially increasing total probability in time. We shall see that these problems can be cured by paying more attention to the auxiliary conditions to define the Hilbert space of pure states, $\cal H$. 

The issue of the boundary conditions is explored from two different angles below. First, in the context of a piecewise constant one-dimensional potential, the non-trivial interactions are represented by the matching conditions, auxiliary conditions for the solutions of the free equation of motion in a restricted spatial region. Such a simple problem already shows the increased importance of the boundary conditions compared to the closed dynamics. The second view on the role of the auxiliary conditions is the identification of the boundary conditions needed to keep the operators occurring in the master equation self-adjoint. This procedure can, in certain cases, replace the missing information of the effective equation of motion method, mentioned above.

\subsection{Piecewise constant potential}
The solutions of the Schr\"odinger equation can easily be obtained for the sum of a piecewise constant and Dirac delta potential. However, the similar, simple extension of the homogeneous open dynamics is not available owing to two kinds of physical complications, the use of a mixed state, and the renormalization of the electric current. We demonstrate the problem by following the traditional strategy, namely, first working out the solution of the equation of motion with constant potential and then matching them at the discontinuities. We set $f=0$ in this calculation for the sake of simplicity.

Let us start with a potential which is the sum of the Dirac delta $\lambda\delta(x)$ and another piecewise constant term with finite discontinuity at $x=0$. The integration through the singularity along $x$ and $x_d$ axes yields the matching conditions
\bea\label{matcond}
\ih\lambda\rho(x,x_d)_{|x=\pm\frac{x_d}2}&=&\pm \mr{Disc}_x J(x,x_d)_{|x=\pm\frac{x_d}2},\nn
2\ih\lambda\rho(x,x_d)_{|x=\pm\frac{x_d}2}&=&\mp\mr{Disc}_{x_d} j(x,x_d)_{|x=\pm\frac{x_d}2},
\eea
in terms of the currents \eq{curr} and \eq{ccurr}. 

In the next step, one solves the master equation with different constant potentials in different coordinate intervals and matches the solutions by Eqs. \eq{matcond}. It is easy to control the Hermiticity and the positivity of the probability density of the coordinate of the full solution; however, to assure the positivity of the density matrix becomes difficult. Hence, one starts with a pure stationary state of the closed dynamics at $t=0$ that turns the open interaction channels on and follows the time evolution of the density matrix for $t>0$. The trivial positivity of the initial pure state is then expected to be preserved due to the Lindblad structure of Eq. \eq{masterl}.

One encounters a difficulty in this procedure for unbounded potential like the Dirac delta, because the first equation of Eqs. \eq{matcond} changes the density matrix in an instantaneous manner along the full coordinate axis $x_\pm$ when the open parameters of the master equations jump to a nonvanishing value. Since one expects that the density matrix remains continuous in time during a sudden turning on of interactions of finite strength, an unbounded potential can not be realized as the limit of a large but finite potential. Hence, we restrict our attention to potentials with finite discontinuity, where the density matrix and its first derivatives remain continuous.

\begin{figure}[ht]
\includegraphics[scale=.3]{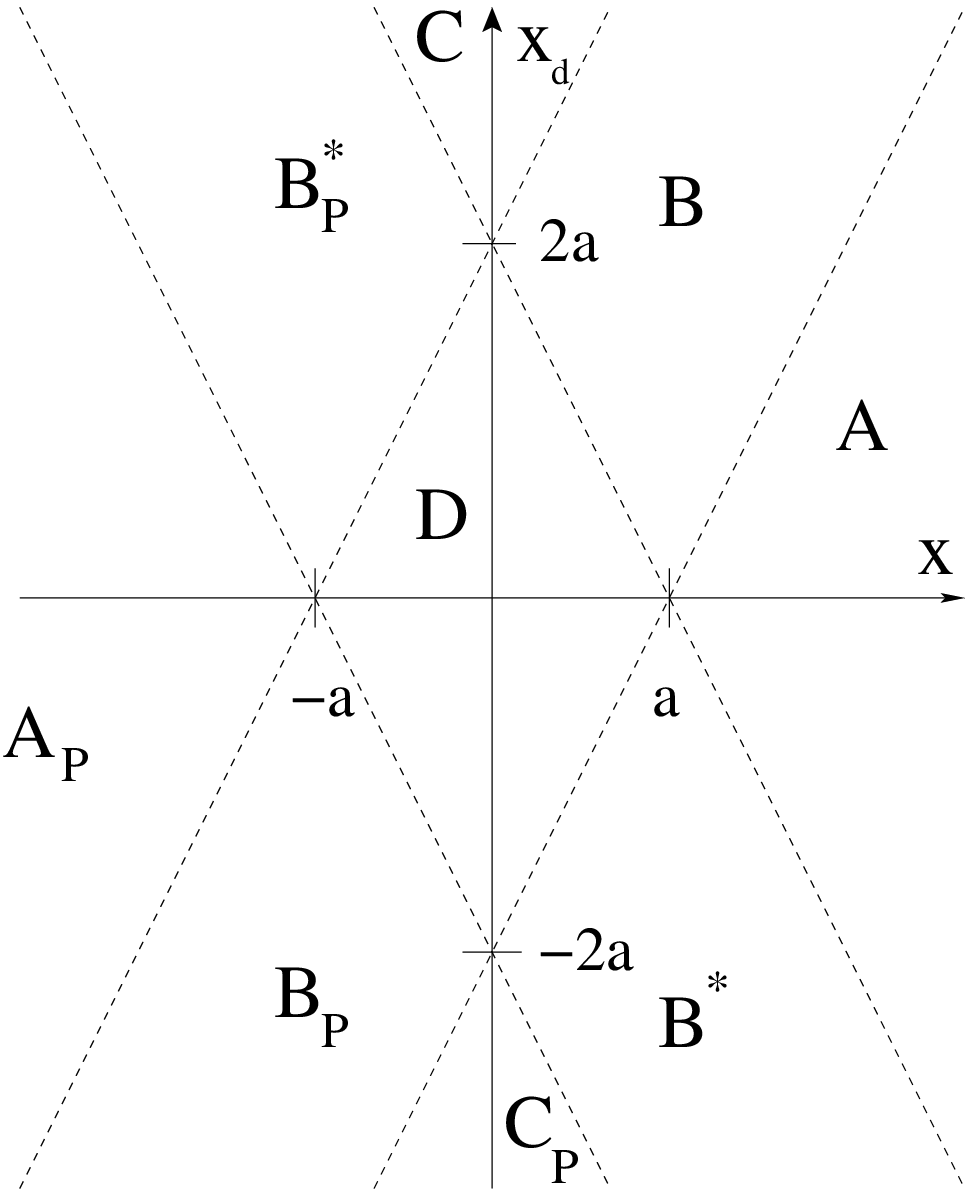}
\caption{The discontinuity of the potential on the plane $(x,x_d)$. The density matrix in the regions with an index $P$ or a star can be obtained by spatial inversion, $x\to-x$, $x_d\to-x_d$, or complex conjugation, respectively.}\label{wellf}
\end{figure}

Let us consider the case of a potential well $U(x)=U_0\Theta(|x|-a)$, when the discontinuities are along the dotted lines in Fig. \ref{wellf} and look for a symmetrical bound state solution $\rho(x,x_d)=\rho(-x,-x_d)$. The initial state at $t=0$ is chosen to be 
\be
\psi(x)=\Theta(|x|-a)e^{-q|x|}+\Theta(a-|x|)c\cos kx,
\ee
where $\hbar k=\sqrt{2mE}$, $\hbar q=\sqrt{2m(U_0-E)}$ with $0<E<U_0$, and $c=(q/k)\cot ka$. The initial density matrix in the regions $A$, $B$, $C$, and $D$ is
\bea\label{initrsp}
\rho_{|A}&=&\rho_{2iq,0},\nn
\rho_{|B}&=&\frac{c}2(\rho_{k+iq,\frac{-k+iq}2}+\rho_{-k+iq,\frac{-k+iq}2}),\nn
\rho_{|C}&=&\rho_{0,iq},\nn
\rho_{|D}&=&\frac{c^2}4(\rho_{2k,0}+\rho_{-2k,0}+\rho_{0,k}+\rho_{0,-k}),
\eea
where $\rho_{q,k}(x,x_d)=e^{iqx+ikx_d}$. The density matrix in the remaining regions is given by $\rho_{|Z_P}(x,x_d)=\rho_{|Z}(-x,-x_d)$, $\rho_{|Z^*}(x,x_d)=\rho^*_{|Z}(x,x_d)$ and $\rho_{|Z^*_P}(x,x_d)=\rho^*_{|Z}(-x,-x_d)$, where $Z=A,B$, or $C$. 

The master equation generates the time dependence \eq{gsol} and \eq{gsolpar} for such a patchwork of initial conditions except $\rho_{|B}\to\rho_{|B}e^{-\ih U_0t}$. However the matching conditions are violated by such a solution. This is easiest to see by noting that the exponential decrease in time of Eq. \eq{gsol} owing to $\Omega_q<0$, the decoherence of the propagating modes, is turned into an exponential increase $\Omega_{iq}>0$ of the tunneling modes and the density matrix becomes discontinuous. In other words, the renormalization of the electric current drives the solution away from the trivial patchwork of homogeneous solutions by generating a nontrivial change around the discontinuity of the potential. The solution of the step function potential, therefore, requires the knowledge of some nontrivial boundary conditions which already contains the fingerprint of the open interaction channels. A possible source of these auxiliary conditions is worked out in the next subsection.

\subsection{Self-adjoint extensions}\label{exts}
We found that the density matrix explodes exponentially in time in a certain region of the $(x,x_d)$ plane and the total probability is not conserved despite the apparent Lindblad structure of the master equation. This problem is due to an actual violation of the Lindblad structure, namely, the momentum operator in Eq. \eq{genmasteq} should be self-adjoint.

The self-adjointness of the momentum operator is a fragile point, and it is assured in the infinite coordinate interval $-\infty<x<\infty$ without boundary by using square integrable wave functions. However one encounters problems when the wave function is either non-normalizable or is sought in a region with boundary, because the momentum-dependent observables are self-adjoint only with a certain extension of their domain of definition realized by appropriate boundary conditions on the wave functions in coordinate space \cite{bonneau}. This problem can be treated in the path integral formalism by the proper definition of  functional space of the paths however, this setting is not transferred by the expansion \eq{expmaeq} to the equation of motion.

The momentum operator is self-adjoint on a finite interval only within the Hilbert space consisting of square integrable wave functions which are periodic up to a fixed phase. While there are boundary conditions which make the square of the momentum operator self-adjoint on a semi-infinite interval, this can not be reached by the momentum operator itself. The situation is slightly different for mixed states. Here, the lack of interference between the additive terms of the density matrix allows that these terms belong to different Hilbert spaces. However, each dyadic product $|\psi\ra\la\psi'|$ must contain bra and kets from the same Hilbert space.

The problems about the self-adjoint extension of the momentum operator can superficially be swept under the rug for the Neumann equation as long as the free Hamiltonian is self-adjoint allowing us to construct a basis set of stationary states. However, the open interactions generate linear terms in the momentum in the master equation \eq{masterlpm} and make a careful treatment of the problem unavoidable. 

To understand the formal origin of the exponential growth of the solution with imaginary $q$, let us suppose that the particle tunnels under a potential in a closed dynamics within a finite coordinate interval where the wave function preserves its phase. The probability flux of the Schr\"odinger equation is vanishing in this case, and the norm of the solution is preserved. The open interactions modify the conserved current by the additive flux 
\be
j_{env}=-\frac{\hbar d_2}{2m^2}e^{-qx}
\ee
which moves more probability inward than outward. Hence, the total probability corresponding to any coordinate interval increases in time due to the lacking self-adjointness of the momentum operator in a coordinate interval with a finite end point.

To recover a physically acceptable dynamics, one has to restrict the pure states into a Hilbert space with self-adjoint extension of the momentum operator. This amounts to the use of the Hilbert space ${\cal H}_\theta=\{\psi(x)|\psi(x+L)=e^{i\theta}\psi(x-L)\}$ in the coordinate interval $-L\le x\le L$ and the density matrices which satisfy the boundary conditions $\rho(x,x_d)=\rho(x+2L,x_d)=e^{-2i\theta}\rho(x,x_d+4L)$. When the regions with nonoscillatory probability density are embedded into the interval $-L<x<L$ the total probability remains time independent. In fact, the total probability is trivially preserved in ${\cal H}_\theta$, where the allowed wave vectors have a discrete spectrum  $q=\pi(n+\theta/2\pi)/L$ with integer $n$. One may make the limit $L\to\infty$ with $\theta=0$ leading to the Hilbert space consisting of square integrable wave functions. The usual textbook examples dealing with the stationary states of piecewise constant potential should be treated by embedding the problem into one such Hilbert space \cite{bonneau}. However, the construction of the proper extension for the master equation remains an open problem.

\section{Summary}\label{concls}
The solutions of the free Schr\"odinger equation is the simplest source of our intuition about closed quantum systems. A similar strategy is followed in this work for open systems by working out few simple solutions of the most general translation invariant harmonic master equation. The summary of the more important conclusions is the following.

\subsection{Kinematics} The space of mixed states is far more rich than in the case of a pure state, namely, the specification of the former requires more information than the latter. This issue comes up in this work in three different contexts. (a) The Neumann equation for the mixed state of a closed dynamics requires more auxiliary conditions than the Schr\"odinger equation. (b) The matching conditions, emerging in the case of piecewise constant potential, are more restrictive for mixed than for pure states. (c) The mixed states of translation invariant dynamics form infinite-dimensional irreducible subspaces as opposed to the one-dimensional irreducible subspaces of the pure states. These complexities arise already in closed dynamics.

\subsection{Effective dynamics} Special care is needed in deriving the effective dynamics in a space with a boundary to assure the self-adjointness of the momentum operator. There are several self-adjoint extensions, and the choice of the actual Hilbert space to represent the physical system is nontrivial. There are two different strategies to follow. When the effective action is sought within the path integral formalism, then the domain of integration, the functional space of the trajectories, has to be carefully defined. When the master equation is sought, then the spatial boundary conditions have to be specified.

\subsection{Electric current} The environment contributes to the conserved electric current by generating form factors to the observed particle. Thereby, the observation of the motion of a charged particle reveals the presence of an environment even if the latter is electrically neutral. In particular, the asymptotic bremsstrahlung of a quantum Brownian motion is halved.

\subsection{Relaxation} The general solution of the master equation within the space of square integrable wave functions is obtained. The dynamics relaxes to a translation invariant mixed state corresponding to the canonical Gibbs density matrix providing a dynamical example of the eigenstate thermalization scenario \cite{deutsch,srednicki,tasaki,calabrese,riemanna,riemannb,linden,riemannc,cho,mueller,palma,gogolin,magan}. This remains valid for infinitesimal system-environment interactions reflecting the noncommutativity of the weak coupling and the long time limits.
 
The approach followed in this work opens several further questions we mention some of them only. How to generalize the patchwork solution of the master equation for piecewise constant potential? A bound state in an attractive potential of a closed dynamics can be viewed as the result of the balance between two coherent processes, the spread of a wave packet and the reflection of the  plane waves on the binding potential. But the coherence is reduced in the presence of open interaction channels. To what extent is the coherent structure of a bound state modified by decoherence? Are there other effects of infinitesimal system-environment interactions than the generation of thermal fluctuations? How do non-Markovian memory effects of the time evolution of the reduced density matrix influence the electric current?

\appendix
\section{Effective equation of motion}\label{peems}
A local Markovian dynamics is described by an effective Lagrangian, which,in turn, can be used to derive the equation of motion for the reduced density matrix. These steps are briefly reviewed in this appendix for a test particle.

\subsection{CTP formalism}\label{ctps}
The observed system and its environment are supposed to be described by the coordinates $\v{x}$ and $\v{y}$ respectively, and we assume that the closed dynamics for the full system, including the observed subsystem and the environment, is defined in the path integral formalism by the help of the action $S[\v{x},\v{y}]=S_s[\v{x}]+S_e[\v{x},\v{y}]$. The time evolution of the density matrix is given by the path integral expression
\bea\label{fable}
\rho(\v{x}_+,\v{y}_+,\v{x}_-,\v{y}_-,t_f)&=&\left(\frac{\sqrt{m_sm_e}}{2\pi i\dt\hbar}\right)^{3N+3}\prod_{n=0}^N\int d^3\hx_nd^3\hy_n\nn
&&\times e^{\ih S[\v{x}_+,\v{y}_+]-\ih S^*[\v{x}_-,\v{y}_-]}\rho(\v{x}_{+,0},\v{y}_{+,0},\v{x}_{-,0},\v{y}_{-,0},t_i),
\eea
over the pair of trajectories $\hat{\v{x}}=(\v{x}_+,\v{x}_-)$, and $\hat{\v{y}}=(\v{y}_+,\v{y}_-)$.  This expression can be obtained by following the usual derivation of the path integral formula for the time evolution operator $U$ and $U^\dagger$, appearing in the full density matrix, $\rho=U\rho_iU^\dagger$, by means of the trajectories $\v{x}_+,\v{y}_+$ and $\v{x}_-,\v{y}_-$. This scheme might be called the open time path formalism, because the pair of paths have freely chosen final points.

The time evolution of the full density matrix \eq{fable} preserves the factorizability and can be split into the trivial product of the matrix elements of $U$ and $U^\dagger$. The power of the CTP formalism becomes evident in calculating the reduced density matrix, $\rho_s=\Tr_e\rho$. In fact, the trace operation closes the environment trajectories, $\v{y}^+(t_f)=\v{y}^-(t_f)$, thereby establishing correlations between the system trajectories $\v{x}_\pm(t)$, which encode the system-environment entanglement and dissipative forces \cite{effth}. Let us choose the initial state at time $t_i$ factorizable for the sake of simplicity: $\rho(\v{x}_+,\v{y}_+,\v{x}_-,\v{y}_-,t_i)=\rho_s(\v{x}_+,\v{x}_-,t_i)\rho_e(\v{y}_+,\v{y}_-,t_i)$. The reduced density matrix can be written in the form
 \be\label{reddm}
\rho_s(\v{x}_+,\v{x}_-,t_f)=\left(\frac{m_s}{2\pi i\dt\hbar}\right)^{3N+3}\prod_{n=0}^N\int d^3\hx_ne^{\ih S_{eff}[\hat{\v{x}}]}\rho_s(\v{x}_{+,0},\v{x}_{-,0},t_i),
\ee
involving the effective action, $S_{eff}[\hat{\v{x}}]=S_s[\v{x}_+]-S^*_s[\v{x}_-]+S_{infl}[\hat{\v{x}}]$, and the influence functional \cite{feynman}
\be\label{inflf}
e^{\ih S_{infl}[\hat{\v{x}}]}=\left(\frac{m_e}{2\pi i\dt\hbar}\right)^{3N+3}\prod_{n=0}^N\int d^3\hy_ne^{\ih S_e[\v{x}_+,\v{y}_+]-\ih S^*_e[\v{x}_-,\v{y}_-]}\rho_e(\v{y}_{+,0},\v{y}_{-,0},t_i),
\ee
where the pair of paths have the same final point, $\v{y}_{+,N+1}=\v{y}_{-,N+1}$. This is the closed time path formalism. Note the difference in the handling way of the observed and the nonobserved coordinates; they are part of the open and closed time path schemes, respectively. 

The final time has been identical for the system and its environment in the previous expressions. Such a restriction can be relaxed by exploiting the unitarity of the time evolution of a closed system which amounts to causality, namely, the possibility that the final time can be chosen arbitrarily after the observation:
\bea\label{timeevops}
\la A(t_o)\ra&=&\Tr AU(t_o,t_i)\rho(t_i)U^\dagger(t_o,t_i)\nn
&=&\Tr U(t_f,t_o)AU(t_o,t_i)\rho(t_i)U^\dagger(t_o,t_i)U^\dagger(t_f,t_o),
\eea
with $t_f\ge t_o$. We take $U(t_f,t_o)$ and $U^\dagger(t_f,t_o)$ in this expression as the time evolution operator of the environment in the absence of the system, allowing the set $t_o$ and $t_f$ as the final time in Eqs. \eq{reddm} and \eq{inflf}, respectively. The possibility of having $t_f\ge t_o$ is especially important for dissipative systems where the environment has a continuous spectrum, requiring $t_f\to\infty$ for $t_o<\infty$. 

The distinguishing feature of the CTP formalism is the reduplication of the degrees of freedom and the representation of the classical dissipative forces and the system-environment entanglement as a coupling between the members of the CTP doublets. While this is an unusual procedure in classical physics, its origin is rather obvious in the quantum case: The expectation value of an observable $A$ in a pure state $|\psi\ra$ of unit norm, 
\be\label{expvalue}
\la A\ra=\sum_{nm}\la\psi|n\ra\la n|A|m\ra\la m|\psi\ra,
\ee
$\{|n\ra\}$ being a basis set, contains two {\em independent} sums over the components which represent the quantum fluctuations of the state $|\psi\ra$. Two probability variables $x_1$ and $x_2$ are independent in probability theory if their joint probability distribution is factorizable: $p(x_1,x_2)=p_1(x_1)p_2(x_2)$. Quantum fluctuations are represented by the probability amplitude in quantum mechanics and the quantum fluctuations of the coordinates of two degrees of freedom can be called independent if the joint probability amplitude, the wave function, is factorizable: $\psi(x_1,x_2)=\psi_1(x_1)\psi_2(x_2)$. In a similar manner, the quantum fluctuations within the bra and the ket, treated in the basis $\{|n\ra\}$ with the factorizable distribution $\rho_{mn}=\la m|\psi\ra\la n|\psi\ra^*$ in Eq. \eq{expvalue}, can be called independent. Thus, quantum mechanics represents a pure state by two independent sets of quantum fluctuations, an obvious statement in view of the factorizable density matrix of a pure state. The two sets of fluctuations are time reversed of each other hence, the pure state dynamics can be described exclusively in terms of kets $|\psi\ra$. 

The factorizability of the quantum fluctuations in the bra and the ket sectors is lost in a mixed state and the correlated fluctuations for the bra and the ket require the use of the ket-bra direct product states $|\psi\ra\la\psi'|$ in the density matrix. In a pure state of a closed dynamics the Hermiticity of the initial density matrix relates the fluctuations in the bra and the ket by complex conjugation, time reversal. The interaction with the environment in an open dynamics introduces correlations during the time evolution as well, requiring the simultaneous treatment of the bra and ket fluctuations. 

The comparison of the derivation of the master equation in Section \ref{masters} with the derivation of the Schr\"odinger equation shows that the ``wave function'' for the redoubled coordinate is actually the density matrix. The independent variation of the two CTP trajectory pair gives rise to the semiholonomic forces in the variational principle of classical mechanics \cite{effth}, and the equation of motion makes the two trajectories identical.

The space of the CTP dynamics is larger than the space where the physical motion takes place. The former is defined by the help of the reduplicated variables, and the latter is its subset, selected by the initial conditions. The initial conditions are time reversed of each other, and one members of the CTP doublet, $x^-$ (bra) in the classical (quantum) case, is time reversed, as well; hence, the physical space contains each degree of freedom only once. This is reminiscent of gauge theory, where the physical states cover a subspace of the full Fock space, and some insight into the dynamics can be gained in both cases by regarding the full, nonphysical space.

\subsection{Effective equation of motion}\label{effeoms}
It is worthwhile to recall that the Schr\"odinger equation appears as a ``Fokker-Planck equation'' in the framework of the path integral formalism. In fact, the Lagrangian \eq{syslagr} yields the time evolution for the wave function:
\be\label{spint}
\psi(\v{x},t)=\left(\frac{m}{2\pi i\dt\hbar}\right)^{\frac32(N+1)}\prod_{n=0}^N\int dx_ne^{\ih\dt\sum_n[\frac{m}2(\frac{\v{x}_{n+1}-\v{x}_n}2)^2-U(\v{x}_{n+1})+\frac{\v{x}_{n+1}-\v{x}_n}\dt\v{a}(\frac{\v{x}_{n+1}+\v{x}_n}2)]}\psi(\v{x}_0,t_i)
\ee
where the scalar potential is $U=e\phi$, the vector potential $\v{a}=e\v{A}/c$ is used with the midpoint prescription and $\v{x}=\v{x}_{N+1}$, $\dt=(t-t_i)/(N+1)$ as $N\to\infty$. The derivation of a local differential equation in time needs the last time step only \cite{schulman}:
\be\label{last}
\psi(\v{x},t+\dt)=\left(\frac{m}{2\pi i\dt\hbar}\right)^{\frac32}\int d^3x'e^{\frac{im}{2\hbar\dt}(\v{x}-\v{x}')^2-i\frac{\dt}\hbar U(\v{x})+\ih(\v{x}-\v{x}')\v{a}(\frac{\v{x}+\v{x}'}2)}\psi(\v{x}',t),
\ee
with a small jump in space, $\v{u}=\v{x}-\v{x}'=\ord{\sqrt{\dt}}$. The expansion in $u$ up to $\ord{\dt}$,
\bea\label{purej}
\psi(\v{x},t+\dt)&=&\left(\frac{m}{2\pi i\dt\hbar}\right)^{\frac32}\int d^3ue^{\frac{im}{2\hbar\dt}
u_j(\delta_{jk}-\frac{\dt}m\partial_ka_j)y_k}\nn
&&\left[1-\ih U(\v{x})+\ih\v{u}\v{a}(\v{x})-\frac1{2\hbar^2}[\v{y}\v{a}(\v{x})]^2\right]\left[1-\v{y}\v{\partial}+\hf(\v{y}\v{\partial})^2\right]\psi(\v{x},t)\nn
&=&\left[1+\frac{\dt}{2m}\v{\partial}\v{a}+i\frac{\hbar\dt}{2m}\Delta-\ih\dt U
+\frac{\dt}{m}\v{a}\v{\partial}-i\frac{\dt}{2m\hbar}\v{a}^2\right]\psi(\v{x},t),
\eea
reproduces
\be\label{fockpl}
i\hbar\partial_t\psi=\left[-\frac{\hbar^2}{2m}\left(\v{\partial}-\ih\v{a}\right)^2+U\right]\psi.
\ee
The expression \eq{spint} is to be compared with a particular Markov process in imaginary time, and Eq. \eq{last} describes the last time step. The leading term of the expansion for small jump $\v{u}$ which gives the Fokker-Planck equation for Markov processes is \eq{purej} and \eq{fockpl} in the present case.

The generalization of the derivation of the equation of motion for the density matrix is straightforward. Since the derivative coupling to the vector potential and in the friction terms have different origins, one should allow different types of point splitting in these terms. Thus, we introduce the dimensionless parameters of the regularization $\eta$, $\eta_\nu$, $\eta_d$, $\eta_\xi$ and write the sum of the Lagrangian \eq{syslagrctp} and \eq{inflagr} with $c_1=c_2=c_3=0$ for two consecutive coordinates $\v{x}(t+\dt)=\v{x}$ and $\v{x}(t)=\v{x}'$ for finite cutoff $\dt$ as
\bea
L&=&m\frac{\Delta\v{x}}{\dt}\frac{\Delta\v{x}_d}{\dt}-m\omega^2\v{x}\v{x}_d-U(\v{x}_+)+U(\v{x}_-)-m\nu\v{x}_d^{(\eta_\nu)}\frac{\Delta\v{x}}{\dt}-m\nu_d\v{x}^{(\eta_d)}\frac{\Delta\v{x}_d}{\dt}\nn
&&+\left(\frac{\Delta\v{x}}\dt+\hf\frac{\Delta\v{x}_d}{\dt}\right)\v{a}\left(\v{x}^{(\eta)}+\frac{\v{x}^{(\eta)}_d}2\right)-\left(\frac{\Delta\v{x}}\dt-\hf\frac{\Delta\v{x}_d}{\dt}\right)\v{a}\left(\v{x}^{(\eta)}-\frac{\v{x}^{(\eta)}_d}2\right)\nn
&&-im\xi\v{x}^{(\eta)}_d\frac{\Delta\v{x}_d}{\dt}+i\frac{d_0}2\v{x}^{d2}+i\frac{d_2}2\left(\frac{\Delta\v{x}_d}{\dt}\right)^2
\eea
by the help of the intermediate points $\v{x}^{(\eta)}=(1-\eta)\v{x}+\eta\v{x}'$. The use of the notation $f_\pm=f(\v{x}_\pm)$, $f=(f_++f_-)/2$, $f_d=f_+-f_-$, and $\v{\nabla}_\pm=\partial/\partial\v{x}_\pm$ leads to the single time step evolution for the reduced density matrix:
\bea\label{expmaeq}
\rho(\hv{x},t+\dt)&=&\left(\frac{m}{2\pi\dt\hbar}\right)^3e^{-\frac{\dt}\hbar(im\omega^2\v{x}\v{x}_d+\frac{d_0}2\v{x}^{d2}+iU_d)}\int d^3yd^3y_d\exp\biggl\{\ih\biggl[\frac{m}\dt\v{y}\v{y}^d\nn
&&+m\nu(\v{x}^d+\eta_\nu\v{y}^d)\v{y}+m\nu_d(\v{x}^d+\eta_d\v{y}^d)\v{y}+im\xi(\v{x}^d+\eta_\xi\v{y}^d)\v{y}^d+i\frac{d_2}{2\dt}\v{y}^{d2}\nn
&&\hskip-13pt-\left(\v{y}+\frac{\v{y}_d}2\right)\v{a}\left(\v{x}+\eta\v{y}+\frac{\v{x}_d+\eta\v{y}_d}2\right)+\left(\v{y}-\frac{\v{y}_d}2\right)\v{a}\left(\v{x}+\eta\v{y}-\frac{\v{x}_d+\eta\v{y}_d}2\right)\biggr]\biggr\}\nn
&&\times\left[1+\sum_{\sigma=\pm}\v{y}_\sigma\v{\nabla}_\sigma+\hf\left(\sum_{\sigma=\pm}\v{y}_\sigma\v{\nabla}_\sigma\right)^2\right]\rho(\hat{\v{x}},t).
\eea
The Gaussian integral is easy to carry out, yielding
\bea
\nabla_{Ut}\rho&=&\left\{i\frac\hbar{m}\v{\nabla}_{Ax}\v{\nabla}_{Ad}+\left[\left(\frac{d_2}m\nu-\xi\right)i\v{x}^d-\nu_d\v{x}\right]\v{\nabla}_{Ax}-\nu\v{x}^d\v{\nabla}_{Ad}+\frac{\hbar d_2}{2m^2}\Delta_{Ax}+{\cal U}\right\}\rho\nn
{\cal U}&=&-\frac{d_0+d_2\nu^2-2m\nu\xi}{2\hbar}\v{x}^{d2}-\ih m(\omega^2+\nu\nu_d)\v{x}^d\v{x}\nn
&&-\eta_d\nu_d-\eta_\nu\nu+\left(1-\frac\eta2\right)\left(i\frac{d_2}{2m^2}\v{\nabla}_x\v{a}_d-\frac1m\v{\nabla}_x\v{a}\right)
\eea
in the continuum limit $\dt\to0$, where the covariant derivatives are defined by Eqs. \eq{covdd}. Gauge invariance requires the suppression of the last term, leading to the midpoint prescription for the gauge field $\eta=1/2$, and the conservation of the total probability imposes the condition $\eta_\nu=\eta_d=0$, leaving $\eta_\xi$ arbitrary, and, finally, translation invariance of the environment sets $\omega=\nu_d=0$, resulting Eq. \eq{master}.

\subsection{Ideal gas environment}\label{iges}
In the case of a large, many-body environment, the coordinate $\v{y}$ is replaced by a field, and it is easy to find the perturbative series for the influence functional by expanding in the system-environment and the internal environment interactions. The former is spelled out in Ref. \cite{gas}, where the effective action of a test particle interacting with a nonrelativistic ideal gas is considered, and the latter can be taken into account by the standard perturbation expansion within the environment. Such a perturbative scheme is especially well suited to a large environment which remains in equilibrium during the interaction with a small system.

The general argument leading to the effective Lagrangian \eq{efflagr} can be realized in a simple model, namely, a test particle moving in an arbitrary potential $U(\v{x})$ and interacting with an ideal gas,
\bea\label{tpiga}
S_s[\v{x}]&=&\int dt\left[\frac{m}2\dot{\v{x}}^2(t)-U(\v{x})\right]\nn
S_e[\v{x},\psi^\dagger,\psi]&=&\int dtd^3y\psi^\dagger(t,\v{y})\left[i\hbar\partial_t+\frac{\hbar^2}{2m}\Delta_y+\mu+V(\v{y}-\v{x}(t))\right]\psi(t,\v{y}),
\eea
where $\mu$ denotes the chemical potential, and the static potential
\be
V(\v{x})=\int\frac{d^3q}{(2\pi)^3}e^{i\v{q}\v{x}}V_{|\v{q}|}
\ee
describes the test particle-gas interaction. The $\ord{V}$ and $\ord{\partial_t^2}$ influence functional yields Eq. \eq{efflagr} \cite{gas} with
\bea\label{effpar}
\delta m&=&\frac1{12\pi^2}\int_0^\infty dqq^4V_q^2\partial_{i\omega}^2D^n_{0,q}\nn
k&=&=-\frac1{6\pi^2}\int_0^\infty dqq^4V_q^4\partial_{i\omega}D^f_{0,q}\nn
d_0&=&-\frac1{6\pi^2}\int_0^\infty dqq^4V_q^2D^i_{0,q}\nn
d_2&=&\frac1{12\pi^2}\int_0^\infty dqq^4V_q^2\partial_{i\omega}^2D^i_{0,q},
\eea
in terms of the propagator $D_{\sigma,\sigma'}(x,y)=-i\hbar\Tr[\psi_\sigma(x)\psi^\dagger_{\sigma'}(y)\rho]$ of the gas particles, yielding $D^n_{\omega,\v{k}}=P1/(\omega-\epsilon_\v{k})$ where $P$ stands for the principal value prescription, $D^f_{\omega,\v{k}}=-i\pi\delta(\omega-\epsilon_\v{k})$, and $D^i_{\omega,\v{k}}=-i\pi\delta(\omega-\epsilon_\v{k})(1+2\xi n_\v{k})$ in the momentum space, where $\epsilon_\v{k}=\hbar^2\v{k}^2/2m_g$, $n_\v{k}=\xi/(e^{\beta(\epsilon_\v{k}-\mu)}-\xi)$ for a gas of temperature $T_e=1/k_B\beta$ and exchange statistics $\xi=\pm1$. 

The quantum-fluctuation-induced temperature \cite{gaussian} reduces to $k_BT_q=\hbar d_0/2m\nu$ in the universal, infinitesimal system-environment interaction limit. The ratio
\be
\frac{\hbar d_0}{2m\nu}=\frac{\hbar\int_0^\infty dqq^4|V_q|^2G^i_{0q}}{2\int_0^\infty dqq^4|V_q|^2\partial_{i\omega}G^f_{0q}}
\ee
can easily be calculated:
\be
\frac{\hbar d_0}{2m\nu}=\frac{\frac{m^2}{\pi\hbar^3\beta}\int_0^\infty dq|V_q|^2\frac{q^3}{1+e^{\beta(\frac{\hbar^2q^2}{8m}-\mu)}}}{\frac{m^2}{\pi\hbar^3}\int_0^\infty dq|V_q|^2\frac{q^3}{1+e^{\beta(\frac{\hbar^2q^2}{8m}-\mu)}}},
\ee
and indicates $k_BT_q=1/\beta$. We find thereby a thermal equilibrium between the relaxed state of the test particle and the infinitely large environment, providing a simple, generic model of thermalization in closed quantum systems.

\acknowledgments
J. P. thanks J\'anos Hajdu for several illuminating discussions.


\begin{thebibliography}{99}
\bibitem{nakajima} S. Nakajima, \journal{Progr. Theor. Phys.}{20}{948}{1958}.
\bibitem{zwanzig} R. Zwanzig, \journal{J. Chem. Phys.}{33}{1338}{1960}.
\bibitem{argyres} P. N. Argyres,  P. L. Kelley, \journal{Phys. Rev.}{134}{A98}{1964}.
\bibitem{grabert} H. Grabert, P. Talkner, P. Hanggi, \journal{Z. Physik}{B26}{389}{1977}.
\bibitem{zeh} E. Joos, H. D. Zeh, \journal{Z. Phys.}{B59}{223}{1985}.
\bibitem{gallis} M. R. Gallis, G. N. Fleming, \journal{Phys. Rev.}{A42}{38}{1990}.
\bibitem{diosi} L. Diosi, \journal{Europhys. Lett.}{30}{63}{1995}.
\bibitem{adler} S. L. Adler, \journal{J. Phys.}{A39}{14067}{2006}.
\bibitem{vacchinie} B. Vacchini, \journal{Phys. Rev.}{E63}{066115}{2001}.
\bibitem{lanz} L. Lanz, B. Vacchini, \journal{Phys. Rev.}{A56}{4826}{1997}.
\bibitem{vacchini} B. Vacchini, \journal{Phys. Rev. Lett.}{84}{1374}{2000}.
\bibitem{hornbergerk} K. Hornberger, \journal{Phys. Rev. Lett.}{97}{060601}{2006}.
\bibitem{dodd} P. J. Dodd, J. J. Halliwell, \journal{Phys. Rev.}{D67}{105018}{2003}.
\bibitem{agarwal} G. S. Agarwal, \journal{Phys. Rev.}{A4}{739}{1971}.
\bibitem{calderialeggett} A. O. Caldeira, A. J. Leggett, \journal{Phys. Rev. Lett.}{46}{211}{1981}; \journal{Phys. Rev. Lett.}{48}{1571}{1982}; \journal{Ann. Phys. (N.Y.)}{149}{374}{1983}.
\bibitem{unruh} W. G. Unruh, W. H. Zurek, \journal{Phys. Rev.}{D40}{1071}{1989}.
\bibitem{feynman}  R. P. Feynman, F. L. Vernon, \journal{Ann. Phys.}{24}{118-173}{1963}.
\bibitem{schw} J. Schwinger, \journal{J. Math. Phys.}{2}{407}{1961}.
\bibitem{keldysh} L. V. Keldysh, \journal{Zh. Eksp. Teor. Fiz.}{47}{1515}{1964}
(\journal{Sov. Phys. JETP}{20}{1018}{1965}).
\bibitem{kamenev} A. Kamenev, {\em Field Theory of Non-Equilibrium Systems} (Cambridge University Press, Cambridge, England, 2011).
\bibitem{rammer} J. Rammer, {\em Quantum Field Theory of Non-Equilibrium States}, (Cambridge University Press, Cambridge, England, 2007).
\bibitem{calzetta} E. A. Calzetta, B. L. A. Hu, {\em Nonequilibrium Quantum Field Theory}, (Cambridge University Press, Cambridge, England, 2008).
\bibitem{hu9293} B. L. Hu, J. P. Paz, Y. Zhang, \journal{Phys. Rev.}{D45}{2843}{1992}; \journal{Phys. Rev.}{D47}{1576}{1993}.
\bibitem{joos} E. Joos in {\em Decoherence and the Appearance of a Classical World in Quantum Theory}, edited by. E. Joos et al. (Springer, New York, 2003).
\bibitem{paz} J. P. Paz, S. Habib, W. H. Zurek, \journal{Phys. Rev.}{D47}{488}{1993}.
\bibitem{macr} J. Polonyi, \journal{Universe}{7}{315}{2021}.
\bibitem{elementary} J. Polonyi, I. Rachid, \journal{Symmetry}{13}{1624}{2021}.
\bibitem{lindblad} G. Lindblad, \journal{Comm. Math. Phys.}{48}{119}{1976}.
\bibitem{sandulescu} A. Sandulescu, H. Scutaru, \journal{Ann. Phys. (N. Y.)}{173}{277}{1987}.
\bibitem{effth} J. Polonyi, \journal{Phys. Rev.}{D90}{065010}{2014}.
\bibitem{peskin} M. E. Peskin, D. V. Schroeder, {\em An Introduction to Quantum Field Theory} (Addison-Wesley, Reading, MA, 1995).
\bibitem{collins} J. C. Collins, A. V. Manohar, M. B. Wise, \journal{Phys. Rev.}{D73}{105019}{2006}.
\bibitem{hu} J. Hu, F. J. Jiang, B. C. Tiburzi, \journal{Phys. Lett.}{B653}{350}{2007}.
\bibitem{irr} J. Polonyi, \journal{Symmetry}{8}{25}{2016}.
\bibitem{gaussian} J. Polonyi, arXiv:1510.03212.
\bibitem{dyndec} J. Polonyi, \journal{J. of Phys.}{A51}{145302}{2018}.
\bibitem{branson} D. Branson, \journal{Am. J. Phys.}{47}{1000}{1979}.
\bibitem{bonneau} G. Bonneau, J. Faraut, G. Valent, \journal{Am. J. Phys.}{69}{322}{2001}.
\bibitem{deutsch} J. M. Deutsch, \journal{Phys. Rev.}{A43}{2046}{1991}.
\bibitem{srednicki} M. Srednicki, \journal{Phys. Rev.}{E50}{888}{1994}.
\bibitem{tasaki} H. Tasaki, \journal{Phys. Rev. Lett.}{80}{1373}{1998}.
\bibitem{calabrese} P. Calabrese, J. Cardy, \journal{Phys. Rev. Lett.}{96}{136801}{2006}.
\bibitem{riemanna} P. Riemann, \journal{Phys. Rev. Lett.}{99}{160404}{2007}.
\bibitem{riemannb} P. Riemann, \journal{Phys. Rev. Lett.}{101}{190403}{2008}.
\bibitem{linden} N. Linden, S. Popescu, A. J. Short, A. Winter, \journal{Phys. Rev.}{E79}{061103}{2009}.
\bibitem{riemannc} P. Riemann, \journal{New J. Phys.}{12}{055027}{2010}.
\bibitem{cho} J. Cho, M. S. Kim, \journal{Phys. Rev. Lett.}{104}{170402}{2010}.
\bibitem{mueller} M. P. M\"uller, E. Adlam, L. Masanes, N. Wiebe,\journal{Comm. Math. Phys.}{340}{499}{2015}.
\bibitem{palma} G. De Palma, A. Serafini, V. Giovannetti, M. Cramer, \journal{Phys. Rev. Lett.}{115}{220401}{2015}.
\bibitem{gogolin} C. Gogolin, J. Eisert, \journal{Rep. Prog. Phys.}{79}{056001}{2016}.
\bibitem{magan} J. M. Mag\'an, \journal{Phys. Rev. Lett.}{116}{030401}{2016}.
\bibitem{schulman} L. S. Schulman, {\em Techniques and Applications of Path Integration} (John Wiley \& Sons, New York, 1981).
\bibitem{gas} J. Polonyi, \journal{Phys. Rev.}{A92}{042111}{2015}.
\end{thebibliography}
\end{document}